\documentclass[njp,12pt]{iopart}
\usepackage{graphicx}
\expandafter\let\csname equation*\endcsname\relax
\expandafter\let\csname endequation*\endcsname\relax
\usepackage{amsmath}
\usepackage{amsfonts}
\usepackage{amssymb}
\usepackage{soul}
\usepackage{dsfont}
\usepackage{hyperref}
\usepackage{amstext}
\usepackage{braket}
\usepackage[caption=false]{subfig}
\usepackage{braket}
\usepackage{color,soul} 
\usepackage{dcolumn}   
\usepackage{bm}        
\usepackage{tabularx,ragged2e,booktabs,caption}
\usepackage{cellspace}
\renewcommand\vec{\boldsymbol}
\usepackage{floatrow} 
\usepackage{hyperref}
\hypersetup{colorlinks=true, citecolor=blue, urlcolor=blue, linkcolor=blue}
\usepackage{color,soul}
\usepackage[a4paper, total={6.5in, 9in}]{geometry}

\date{\today}

\begin{document}

\title{Mesoscopic entanglement through central-potential interactions}

\author{Sofia Qvarfort$^{1,2}$, Sougato Bose$^2$, Alessio Serafini$^2$}
\address{$^1$ QOLS, Blackett Laboratory, Imperial College London,  SW7 2AZ London, United Kingdom \\
$^2$ Department of Physics and Astronomy, University College London, Gower Street, WC1E 6BT London, United Kingdom }
\ead{sofiaqvarfort@gmail.com}

\begin{abstract}
The generation and detection of entanglement between mesoscopic systems would have major fundamental and applicative implications. In this work, we demonstrate the utility of continuous variable tools to evaluate the Gaussian entanglement arising between two homogeneous levitated nanobeads interacting through a central potential. We compute the entanglement for the steady state and determine the measurement precision required to detect the entanglement in the laboratory. 
\end{abstract}

\section{Introduction}
The mastery over levitated optomechanical systems in the laboratory is becoming increasingly refined. With advances in cooling a small levitated sphere to the ground-state~\cite{teufel2011sideband, chan2011laser,millen2015cavity, jain2016direct, delic2020cooling} and 
in the preparation of squeezed states~\cite{rashid2016experimental}, compounded by the ability to control and implant charges into levitated nanobeads~\cite{neukirch2015multi}, setups are reaching unprecedented levels of control. Furthermore, optomechanical systems have shown significant potential for sensing applications~\cite{gavartin2012hybrid, xu2014squeezing}, especially with regards to measuring gravitational parameters~\cite{qvarfort2018gravimetry, armata2017quantum, marshman2018mesoscopic}.  As a result, the  engineering of schemes to entangle multiple levitated oscillators and, more broadly, mesoscopic systems, is being identified as a major medium-term milestone of the field~\cite{nagata2007beating}. In fact, the entanglement for a number of mesoscopic systems has already been demonstrated experimentally~\cite{jost2009entangled, appel2009mesoscopic, riedinger2018remote}, although a more thorough and exhaustive 
understanding of the conditions under which entanglement may be generated is required to move on to applications. 

A key property of entanglement is that it acts as an unambiguous hallmark of non-classicality. As a result, entanglement between mesoscopic systems can aid the quest of mapping the transition from the quantum to the classical scale~\cite{paternostro2007creating, milman2005proposal}. Furthermore, recent proposals concerning the fundamental nature of gravity have considered quantum entanglement as generated by a Newtonian potential between two massive quantum systems~\cite{bose2017spin, marletto2017gravitationally}. The Newtonian potential can be seen as an effective potential that arises form an underlying fully quantum field theory.  Successfully detecting gravitational entanglement would indeed be a strong indication of the quantisation of gravity~\cite{belenchia2018quantum} and, in general, the detection of entanglement as generated by gravity has significant ramifications, in the attempt to feed theories of quantum gravity with new empirical evidence at low energies~\cite{carney2018massive}.

The experimental setup  envisioned in~\cite{bose2017spin} is challenging since it requires the generation of highly localised spatial superpositions with large separation for two levitated systems. Gaussian states, on the other hand, are more well-understood, and they can be straight-forwardly prepared in the laboratory. As such, we wish to explore whether the questions posed in~\cite{bose2017spin} can be tested with Gaussian states only. Furthermore, to detect entanglement from the Newtonian potential, it is reasonable to first consider an analogous case with a stronger potential, such as the Coulomb potential. 

Our main question in this work is therefore: Is it possible to detect entanglement between two mesosopic systems as generated by a  central potential interaction with only Gaussian resources? 
To address this question, we consider a fundamental or effective central potential of the form $1/r^n$, for integer $n$ and where $r = |\vec{r}_1 - \vec{r}_2|$ with position vectors $\vec{r}_1$ and $\vec{r}_2$ acting between two spherically symmetric quantum systems. Crucially, we ask whether minimal initial state preparation, such as squeezing, as opposed to severe requirements of preparing  highly non-Gaussian states (as, for example, proposed in~\cite{bose2017spin}) is enough to generate detectable entanglement, and whether the witnessing of the generated entanglement is possible simply by measuring position--momentum correlations. We then quantify the leading-order contribution to entanglement within the continuous-variable (CV) framework for both the dynamical generation of entanglement from initial squeezed states of the interacting oscillators and for the system's steady-state in the presence of noise.

\begin{figure*}
  \includegraphics[width=0.8\linewidth]{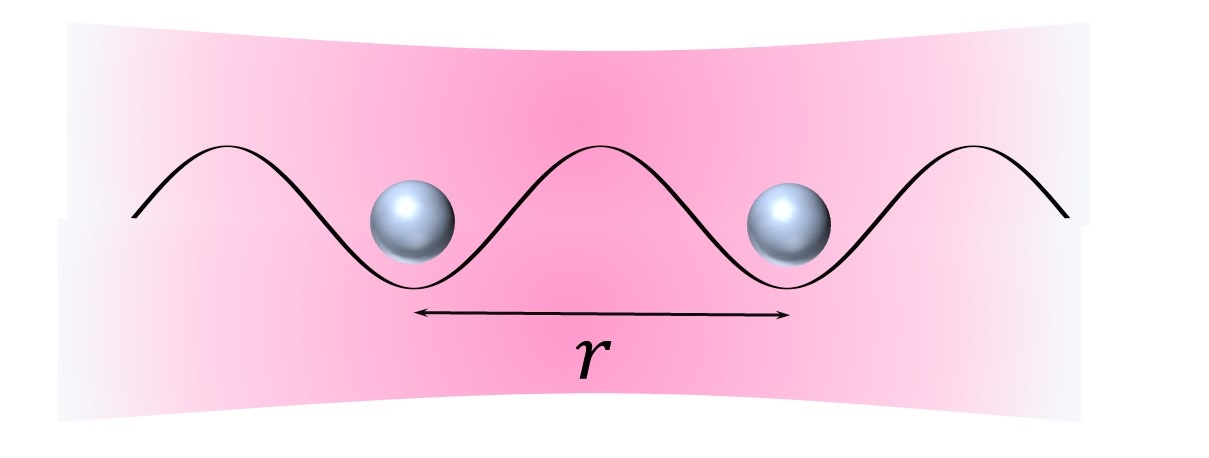}%
\caption{\small Two nanobeads in a laser trap suspended a distance $r$ apart. The beads are allowed to interact via a central potential, such as the Coulomb potential or the Newtonian potential.}
\label{fig:spheres}
\end{figure*}


\section{Dynamics}
We begin by considering two spheres trapped next to each other as per Figure~\ref{fig:spheres}. The non-interacting system Hamiltonian $\hat{H}_0$  that describes the harmonic motion for both spheres in the local trap is given by 
\begin{equation}
\hat{H}_0 = \frac{1}{2} m_1 \, \omega_1 ^2 \,  \hat{x}_1^2 + \frac{1}{2} m_2 \,  \omega_2^2 \,  \hat{x}_2^2 + \frac{\hat{p}_1^2}{2m_1} + \frac{\hat{p}_2^2}{2m_2}, 
\end{equation}
where $m_1$ and $m_2$ are the system masses, $\omega_1$ and $\omega_2$ are the respective mechanical trapping frequencies for each sphere, and $\hat{x}_i$ and $\hat{p}_i$ with $i = 1,2$ are the position and momentum operators for system 1 and 2 respectively.  

We now consider a generic central potential of the form $\alpha/|\vec{r}_1 - \vec{r}_2|^n$,  where $\alpha$ is the coupling constant and $\vec{r}_1$ and $\vec{r}_2$ are position vectors. In all cases considered here, this potential is the lowest-order approximation to a description that includes the full field theory. We proceed to derive the interaction from the generic central potential to second order. The quadratic terms will capture the largest contribution to the central-potential entanglement and retain the quadratic interaction, which maps input Gaussian states to output Gaussian states. The last effect allows us to model the system completely within the covariance matrix formalism~\cite{serafini}. 

To derive the Hamiltonian interaction term, we assume that the movement of the spheres is constrained in all  but the $x$-direction, which is the axis along which the two systems are trapped. We consider small perturbations to $r_1$ and $r_2$, such that $r_1 \rightarrow r_1 - x_1$ and $r_2 \rightarrow r_2 - x_2$, with $x_1 \ll r_1$ and $x_2 \ll r_2$. By denoting $r = r_1 - r_2$ and $\Delta x = x_1 - x_2$, we Taylor-expand the interaction to second order in $\Delta x$ to find:
\begin{align} \label{eq:Taylor}
\frac{1}{(r - \Delta x)^n} &= \frac{1}{r^n} + n\frac{\Delta x }{r^{n+1}} + \frac{n(n+ 1)}{2} \frac{(\Delta x)^2}{r^{n+2} } + \ldots , 
\end{align}
where we have left out the dimensional prefactor $\alpha$ and ignored all terms of order $\mathcal{O}\left[ (\Delta x)^3 \right]$. We then quantise the positions of the two masses around their equilibrium positions, which are taken to be a distance $r$ apart  from each other, by promoting the position coordinates to operators: $x_i \rightarrow \hat{x}_i$. For gravity, this step incorporates the assumption that gravity is a quantum force that can path-entangle two quantum systems~\cite{bose2017spin}.

To arrive at the entangling interaction term, we discard the constant term, which is the first term on the left-hand side in Eq.~\eqref{eq:Taylor}, as its only contribution is a static energy shift. Secondly, since a displacement term in a quadratic Hamiltonian does not affect the entanglement~\cite{serafini}, we can also discard the term linear in position, which is the second term on the left-hand side in Eq.~\eqref{eq:Taylor}. The final term, however, contains a mixing of the position operators in the form $\hat{x}_1 \hat{x}_2$, which will generate entanglement between the two states. The interaction term in the Hamiltonian thus becomes
\begin{align} \label{eq:interaction:Hamiltonian}
\hat{H}_{I} &=\alpha \,  \frac{n(n+1)}{2 \,r^{n+2}} ( \hat{x}_1 - \hat{x}_2)^2 \, .
\end{align}
The full Hamiltonian therefore reads
\begin{equation} \label{eq:full:Hamiltonian}
\hat{H}= \frac{1}{2} m_1 \, \omega_1 ^2 \,  \hat{x}_1^2 + \frac{1}{2} m_2 \,  \omega_2^2 \,  \hat{x}_2^2 + \frac{\hat{p}_1^2}{2m_1} + \frac{\hat{p}_2^2}{2m_2} + \alpha \,  \frac{n(n+1)}{2 \,r^{n+2}} ( \hat{x}_1 - \hat{x}_2)^2. 
\end{equation}
Since $\hat H$ is quadratic in the canonical operators, all initial Gaussian states exclusively evolve into Gaussian states. Furthermore, all Gaussian states are uniquely defined by their first and second moments, which allows us to model this system completely within the covariance matrix framework~\cite{serafini}. We introduce the $4\times4$ two-mode covariance matrix $\boldsymbol{\sigma}$, which consists of all second moments of the Gaussian state $\hat{\rho}_{\rm{G}}(t)$. It is defined as 
\begin{equation}
\boldsymbol{\sigma}(t) = \mathrm{Tr} \left[  \{\hat{\boldsymbol{r}}, \hat{\boldsymbol{r}}^{\mathrm{T}} \} \, \hat \rho_{\rm{G}}(t) \right] \, , 
\end{equation}  
where the vector of operators is given by $\hat{\boldsymbol{r}} = (\hat{x}_1, \, \hat{p}_1,\,  \hat{x}_2, \, \hat{p}_2)^{\mathrm{T}}$, and where the bracket $\{\cdot, \cdot \}$ denotes the symmetrised outer product, in the sense that  $\{\hat{\vec{r}}, \hat{\vec{r}} ^{\rm{T}} \} = \hat{\vec{r}} \hat{\vec{r}}^{\rm{T}} + (\hat{\vec{r}} \hat{\vec{r}}^{\rm{T}})^{\rm{T}}$~\cite{serafini}. 

The evolution of the second moments under the Hamiltonian in Eq.~\eqref{eq:full:Hamiltonian} can be encoded through the symplectic matrix $S = e^{\Omega \, H \, t /\hbar}$, where $H$ is the Hamiltonian matrix, defined as
\begin{equation} \label{eq:def:Hamiltonian:matrix}
\hat H = \frac{1}{2} \hat{\boldsymbol{r}}^{\rm{T}} H \hat{\vec{r}} + \hat{\boldsymbol{r}}^{\rm{T}} \boldsymbol{r} \,,
\end{equation}
and where ${\Omega}$ is the symplectic form defined in this basis as 
\begin{align}
{\Omega} =\bigoplus_{j=1}^n \,  \omega \, ,  \,  \mbox{ with  } \quad &\omega =   \left( \begin{matrix} 0 & 1 \\
- 1 & 0 \end{matrix} \right) \, ,
\end{align}
for a total of $n$ modes. For a bipartite system like the one considered here, $n = 2$, which means that $\Omega$, $\boldsymbol{\sigma}$, $H$, and $S$ are all $4\times 4$ matrices. 
The covariance matrix $\boldsymbol{\sigma}$ then evolves as $\boldsymbol{\sigma}(t) = S \, \boldsymbol{\sigma}_0  \, S^{\mathrm{T}}$, where $\boldsymbol{\sigma}_0 $ encodes the second moments of the initial state.

To determine the Hamiltonian matrix ${H}$ that arises from Eq.~\eqref{eq:full:Hamiltonian}, we define the dimensionless operators $\hat{x}'_i$ and $\hat{p}_i^\prime$ as $\hat{x}_i = \sqrt{\hbar/(m_i \, \omega_{i})}  \, \hat{x}^\prime_i$ and $\hat{p}_i = \sqrt{\hbar \, m_i \, \omega_{i}}  \, \hat{p}^\prime_i$. 
For notational simplicity, we then assume that $m_1 = m_2 = m$, and $\omega_1 = \omega_2 = \omega_{\mathrm{m}}$. The Hamiltonian in Eq.~\eqref{eq:full:Hamiltonian} therefore becomes
\begin{align} \label{eq:rescaled:Hamiltonian}
\hat{H} &=  \frac{\hbar \omega_{\mathrm{m}} }{2} \left( \hat{x}_1^{\prime2}  + \hat{p}_1^{\prime2}\right)   +  \frac{\hbar \omega_{\mathrm{m}}}{2} \left( \hat{x}_2^{\prime2} + \hat{p}^{\prime2}_2 \right) + \frac{\alpha \, \hbar}{\omega_{\mathrm{m}} m}  \frac{n(n+ 1)}{2\, r^{n + 2} }( \hat{x}_1' - \hat{x}_2')^2. 
\end{align}
In what follows, we will rescale the laboratory--time $t$ by $\omega_{\mathrm{m}}$ to obtain the  dimensionless time parameter $\tau = \omega_{\mathrm{m}} t$. We later consider open-system dynamics where $\kappa$ denotes a mechanical decoherence rate. Here, we also rescale $\kappa$ to $\tilde{\kappa} = \kappa/\omega_{\rm{m}}$. 

This yields the Hamiltonian matrix $\tilde{{H}} = {H} /(\hbar \, \omega_{\mathrm{m}})$:
\begin{align} \label{trapped:full:Hamiltonian:matrix:rescaled}
\tilde{{H}} &=  {H}_0 +  \tilde{\alpha} \, {H}_{I}^{(1)} - \tilde{\alpha} \, {H}_{I}^{( 2)}, 
\end{align}
where for convenience we have defined the dimensionless coupling 
\begin{equation} \label{eq:rescaled:coupling}
\tilde{\alpha} = \frac{\alpha  \, n (n  +1)}{ \omega^2_{\mathrm{m}} m \, r^{n+2}},
\end{equation} 
and where ${H}_0 = \mathds{1}_4$ is the $4\times 4 $ identity matrix that governs the free evolution and ${H}_I^{(i)}$ denotes the Hamiltonian matrices responsible for the interaction. They are given by 
\begin{align} \label{eq:trapped:Hamiltonian:matrix}
&{H}_{I}^{(1)} =  \begin{pmatrix} 1 & 0 & 0 & 0 \\ 0 & 0 & 0 & 0 \\ 0 & 0 & 1 & 0 \\ 0 & 0 & 0 & 0 \end{pmatrix} \,, &{H}_I^{(2)} =\begin{pmatrix} 0 & 0 & 1 & 0 \\ 0 & 0 & 0 & 0 \\ 1 & 0 & 0 & 0 \\ 0 & 0 & 0 & 0 \end{pmatrix}  \, ,
\end{align}
where specifically ${H}_I^{(2)}$ will generate the entanglement. We  note that ${H}_I^{(2)}$ is of the form of two-mode squeezing, which implies that the corresponding closed system will not display periodic behaviour.  With these rescaled quantities, the evolution is now encoded as $S = e^{\Omega \,   \tilde{H} \, \tau}$.


\section{Computing the entanglement}
To compute the entanglement that arises from the central-potential interaction, we make use of the logarithmic negativity~\cite{lee2000partial, vidal2002computable, plenio2005logarithmic}, a well-known monotone that quantifies the degree of violation of the positive-partial-transpose (PPT) criterion~\cite{peres1996separability,horodecki1997separability}. The latter is inspired by the by the fact that, given a separable state $\hat \rho = \hat \rho_A \otimes \hat \rho _B$, the partial transpose with respect to one of the subsystems leaves the state with positive eigenvalues: $\hat \rho^{\rm{Tp}} \geq 0$. Hence, should we find that $\hat \rho^{\rm{Tp}} < 0$, the state is entangled. Notice that this criterion turns out to be necessary and sufficient for Gaussian states~\cite{simon2000peres, lami2018gaussian}.

The PPT criterion in the CV framework can be explicitly computed by dividing $\boldsymbol{\sigma} $ into submatrices $\boldsymbol{\sigma}_A, \boldsymbol{\sigma}_B$ and $\boldsymbol{\sigma}_{AB}$ as such~\cite{adesso2014continuous}:
\begin{align}
\boldsymbol{\sigma} = \left( \begin{matrix} \boldsymbol{\sigma}_A & \boldsymbol{\sigma}_{AB} \\ \boldsymbol{\sigma}_{AB} & \boldsymbol{\sigma}_B \end{matrix} \right).
\end{align}
We define the symplectic invariant quantity $\Delta =  \mathrm{det}\, \boldsymbol{\sigma}_A + \mathrm{det}\,\boldsymbol{\sigma}_B + 2\,  \mathrm{det} \,  \boldsymbol{\sigma}_{AB}$. In this basis, the partial transpose is equivalent to setting $\hat{p}_i \rightarrow - \hat{p}_i$ for one subsystem, which implies $\Delta \rightarrow \tilde{\Delta} = \mathrm{det} \,  \boldsymbol{\sigma}_A + \mathrm{det} \, \boldsymbol{\sigma}_B - 2 \,  \mathrm{det} \,  \boldsymbol{\sigma}_{AB}$.  The positive partial transpose (PPT) criterion for two-mode Gaussian states can thus be compactly expressed as $\mathrm{det} \, \boldsymbol{\sigma} - \tilde{\Delta} + 1 \geq 0$~\cite{serafini}. When this equality is violated, the state is entangled. For bipartite states, this is a necessary and sufficient condition for entanglement~\cite{simon2000peres}. 

To quantify the entanglement, we calculate the logarithmic negativity $E_N$ of the state, defined by $E_N ( \boldsymbol{\sigma}) = \max{(0, - \log_2{\tilde{\nu}_- })}$, where $\tilde{\nu}_\mp$ are the symplectic eigenvalues of the state, defined for bipartite systems as 
\begin{equation}
\tilde{\nu}^2_\mp = \frac{\tilde{\Delta} \mp \sqrt{\tilde{\Delta}^2 - 4 \,  \mathrm{det} \, \boldsymbol{\sigma}}}{2} \, .
\end{equation}
In this work, we consider a two-mode mechanically squeezed state $\boldsymbol{\sigma}_S= \mathrm{diag}(z , z^{-1}, z^{-1}, z)$ as the initial state. Note that the squeezing occurs in the opposite quadrature, which serves to increase the entanglement from this particular interaction. When $z = 1$, the state is a coherent state with $\boldsymbol{\sigma}_0 = \mathrm{diag}(1,1,1,1)$. There are many ways to mechanically squeeze the system to produce $\boldsymbol{\sigma}_S$, including controlling the trap frequency or using a Duffing non-linearity~\cite{lu2015steady}. In addition, thermal optomechanical squeezing has been  experimentally realised in~\cite{rashid2016experimental}.

\begin{figure}[t!]
\subfloat[ \label{fig:entanglement:vs:time:various:alpha}]{%
  \includegraphics[width=.45\linewidth,trim = 0.5cm 0cm -0.5cm 0cm]{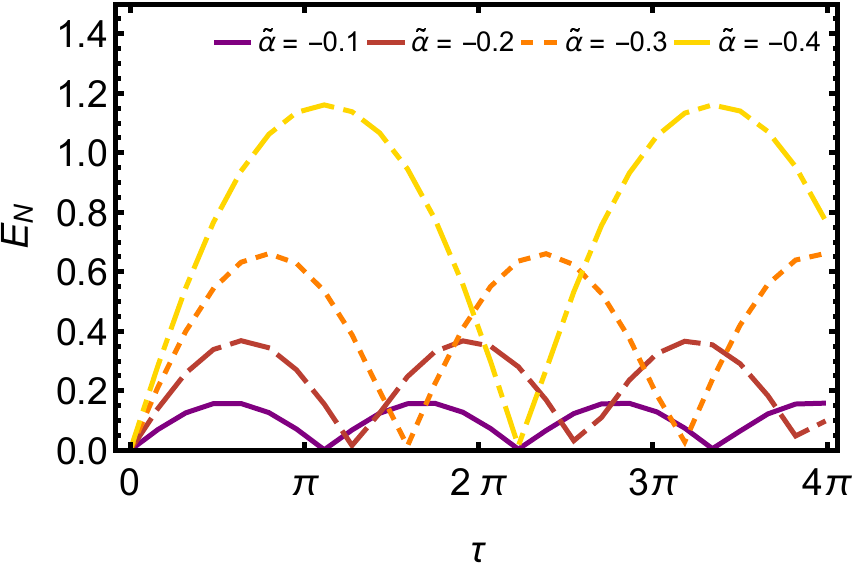}%
}\hfill $\quad$
\subfloat[ \label{fig:entanglement:vs:time:various:z}]{
  \includegraphics[width=.435\linewidth,trim = 0.5cm 0.5mm -0.5cm -0.5mm]{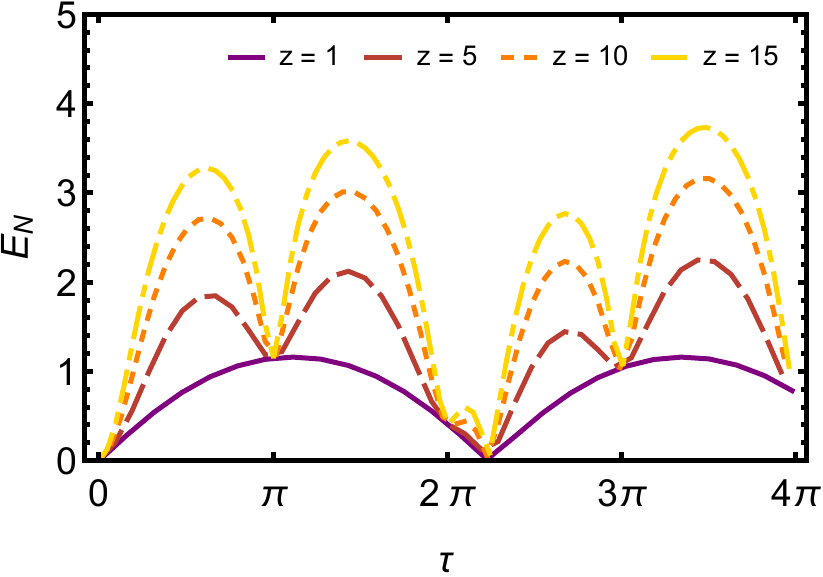}%
}\hfill
\caption{\small Entanglement from central-potential interactions. 
 \textbf{(a)} Plot of the logarithmic negativity $E_N$  as a function of time $\tau$ for squeezing $z = 1$ and different values of the rescaled coupling $\tilde{\alpha}$.  \textbf{(b)} Plot of $E_N$ as a function of time $\tau$ for different squeezing parameters $z$ and $\tilde{\alpha} = -0.4$.  The oscillations are due to the rotating terms contained in the $\hat x_1 \hat x_2$ interaction (note however that $E_N$ never goes fully to zero).}
\label{fig:general:dynamics}
\end{figure}

For the system to be thermally stable (i.e.~for the Hamiltonian to be bounded from below), we require that the rescaled Hamiltonian matrix $\tilde{H}$ in Eq.~\eqref{trapped:full:Hamiltonian:matrix:rescaled} satisfies $\tilde{H}> 0$, which means that the eigenvalues of $\tilde{H}$ must be positive~\cite{serafini}. For $\tilde{H}$ in Eq.~\eqref{trapped:full:Hamiltonian:matrix:rescaled}, we find that the eigenvalues $\lambda_i$ for $i = 1,2,3,4$ are given by $\lambda_{1} = \lambda_2 = \lambda_3 = 1$, and $\lambda_4= 1 + 2 \tilde{\alpha}$. This means that for a positive (repulsive) coupling $\tilde{\alpha}$, there is no limit to the interaction strength that will unbound the Hamiltonian. We are therefore able to obtain considerable amounts of entanglement for a repulsive potential. When $\tilde{\alpha}$ is negative (attractive), however, the maximum strength of the interaction is limited to $\tilde{\alpha} > - 1/2$. Since all three potentials considered in this work can be attractive (with $\tilde{\alpha} <0$), and because strong repulsive potentials pose the risk of eventually losing the trapped beads, we will mainly explore the values that can be obtained for $\tilde{\alpha} \sim -0.4$, which is close to the limiting value. 

We now proceed to compute the entanglement $E_N$. While an analytic expression for $E_N$ is available, it is too long  and cumbersome to reproduce here. Instead, we plot the entanglement $E_N$ for different values of $\tilde{\alpha}$ in Fig.~\ref{fig:general:dynamics}. 
In Fig.~\ref{fig:entanglement:vs:time:various:alpha} we plot $E_N$ for  coherent states with $z = 1$ as a function of time $\tau$ for various values of $\tilde{\alpha}$. The stronger the coupling, the more entanglement is generated.  
Next, in Fig.~\ref{fig:entanglement:vs:time:various:z} we plot the same general dynamics for $\tilde{\alpha} = -0.4$ but for different $z$.  We set $\tilde{\alpha} =- 0.4$ because it is close to the maximally thermally stable value of $\tilde{\alpha} = - 0.5$.  Increasing $z$ serves to amplify the already-present entanglement. 



\section{Open system dynamics}
All systems are subject to environmental noise that generally degrades the entanglement present in the system. In this work, we consider two types of noise: damping of the oscillator motion in terms of phonon decay, which we denote $\tilde{\kappa} = \kappa/ \omega_{\mathrm{m}}$, and the number of thermal phonons $N_{\mathrm{th}}$ present in the system. 

For Markovian dynamics, the  covariance matrix $\boldsymbol{\sigma}(\tau)$ evolves as 
\begin{equation} \label{eq:noisy:evolution}
\dot{\boldsymbol{\sigma}} = {A}(\tau) \, \boldsymbol{\sigma} + \boldsymbol{\sigma} {A}^{\mathrm{T}}(\tau)  + {D}, 
\end{equation}
where ${A}(\tau) = {\Omega} H(\tau) - \tilde{\kappa} \,  \mathbb{I}_4/ 2$ is a drift matrix that incorporates the Hamiltonian matrix $H$ defined in Eq.~\eqref{eq:def:Hamiltonian:matrix}, and ${D} = (2 N_{\mathrm{th}} + 1) \, \tilde{\kappa} \, \mathbb{I}_4$, with $\tilde{\kappa} = \kappa/\omega_{\rm{m}}$ being the rescaled phonon dissipation rate, $N_{\mathrm{th}}$ the number of thermal phonons present in the system and $\mathbb{I}_4$ the $4\times4$ identity matrix. 

\begin{figure*}
\subfloat[ \label{fig:noisy:logneg:vs:time:various:gamma}]{%
  \includegraphics[width=.45\linewidth,trim = 0.49cm 0cm -0.49cm 0.19cm]{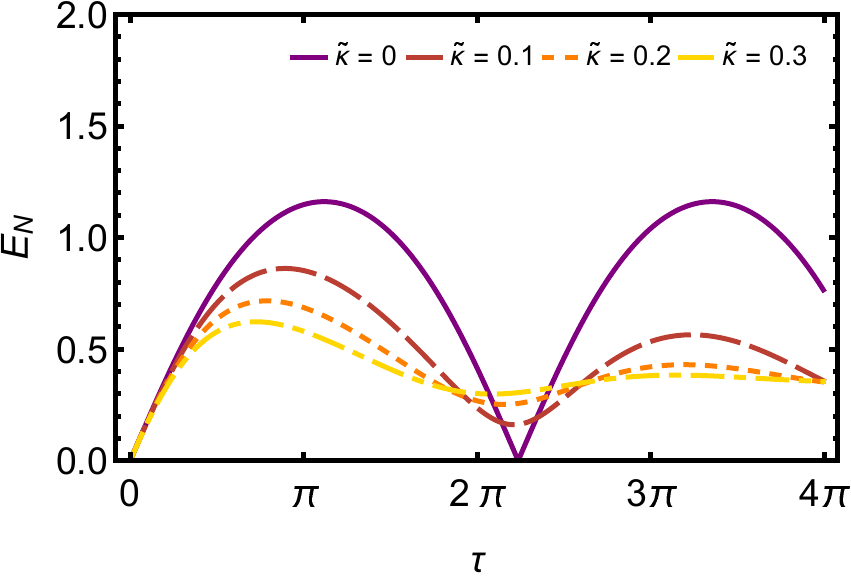}%
}\hfill $\quad$
\subfloat[  \label{fig:noisy:logneg:vs:time:various:z}]{
  \includegraphics[width=.45\linewidth,trim = 0.49cm 0cm -0.49cm 0cm]{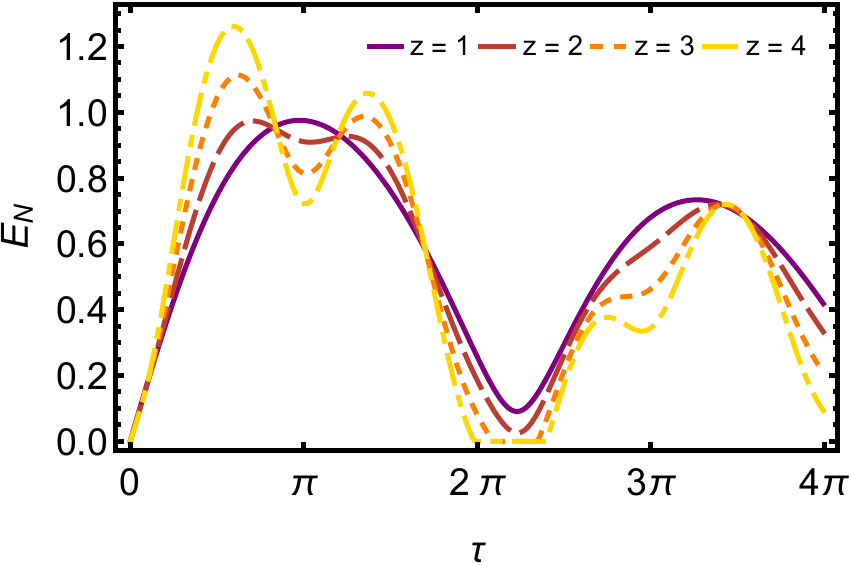}%
}\hfill
\caption{\small Open system entanglement from  central-potential interactions. 
 \textbf{(a)} Plot of $E_N$ as a function of time $\tau$ for increasingly noisy systems with decoherence rate $\tilde{\kappa}$ at $\tilde{\alpha} = -0.4$ and $N_{\rm{th}}$. (\textbf{b}) Plot of $E_N$ as a function of time $\tau$ for different values of squeezing $z$ with decoherence rate $\tilde{\kappa} = 0.05$ at $\tilde{\alpha} = -0.4$ and $N_{\rm{th}} = 0$. }
\label{fig:general:dynamics:noisy}
\end{figure*}

To determine the effect of noise on the entanglement, we evolve the system through Eq.~\eqref{eq:noisy:evolution}. As expected, we  find that the entanglement decreases with time $\tau$ as the systems decoheres. 
We plot the effects of noise on $E_N$ in Fig.~\ref{fig:general:dynamics:noisy}.  In Fig.~\ref{fig:noisy:logneg:vs:time:various:gamma}, we have plotted  $E_N$ as a function of time $\tau$ for a noisy environment for different $\tilde{\kappa}$ at $\tilde{\alpha} = -0.4$,  $z = 1$ and $N_{\rm{th}} = 0$.
Similarly, in Fig.~\ref{fig:noisy:logneg:vs:time:various:z}, we have plotted $E_N$ as a function of rescaled time $\tau$ for different squeezing values $z$ at $\tilde{\alpha} = -0.4$ and $N_{\rm{th}} = 0$. We note that while increasing the squeezing $z$ causes  $E_N$ to increase at first, higher squeezing rates also makes the system more sensitive to noise, a fact that finds extensive confirmation in the existing literature~\cite{serafini2004entanglement, serafini2005quantifying, shackerley2017locally}.  

\subsection{Steady state entanglement}
After long times $\tau \gg 1$ the system enters  a steady state where the entanglement remains at a fixed value. We can obtain this state by solving the matrix equation ${A} \, \boldsymbol{\sigma }^{(\infty)} + \boldsymbol{\sigma}^{(\infty)} {A}^{T} + {D} = 0$  to find the steady state $\boldsymbol{\sigma}^{(\infty)}$, which is defined by the property $\dot{\boldsymbol{\sigma}}^{(\infty)}= 0$. We find the following elements of $\boldsymbol{\sigma}^{(\infty)}$: 
\begin{align} \label{eq:steady:state:covariance:elements}
\sigma_{11}^{(\infty)} &= \frac{ 6 \, \tilde{ \alpha} +\tilde{\kappa }^2+4}{8 \,  \tilde{\alpha} +\tilde{\kappa}^2+4}  \, (2  \, N_{\rm{th}}+1) \, ,\nonumber \\
\sigma_{22}^{(\infty)} &=\frac{4 \, \tilde{\alpha}^2+10 \, \tilde{\alpha} + \tilde{\kappa} ^2+4}{8 \, \tilde{\alpha} +\tilde{\kappa}^2+4} \, (2  \, N_{\rm{th}}+1)  \, ,\nonumber \\
\sigma_{12}^{(\infty)} &=\sigma_{34}^{(\infty)} =    -\frac{\tilde{\alpha } \, \tilde{ \kappa}  }{8  \,  \tilde{\alpha} + \tilde{\kappa}^2+4} \, (2  \, N_{\rm{th}}+1) \, , \nonumber \\
\sigma_{14}^{(\infty)} &= \sigma_{23}^{(\infty)} = \frac{\tilde{\alpha} \, \tilde{\kappa}}{8 \, \tilde{\alpha} + \tilde{\kappa}^2 + 4} \, (2  \, N_{\rm{th}}+1)\, , \nonumber \\
\sigma_{13}^{(\infty)} &= \frac{2 \, \tilde{\alpha}}{ 8 \, \tilde{\alpha} + \tilde{\kappa}^2 + 4} ( 2 \, N_{\rm{th}} + 1)  \, ,\nonumber \\
\sigma_{24}^{(\infty)} &= -\frac{2 \, \tilde{ \alpha}   \, (2  \, \tilde{\alpha} +1) }{8  \, \tilde{\alpha} + \tilde{\kappa }^2+4} \, (2 \, N_{\rm{th}}+1) \, .
\end{align}
All other elements follow by symmetry  from the fact that $(\boldsymbol{\sigma}^{(\infty)})^{ \mathrm{T}} = \boldsymbol{\sigma}^{(\infty)}$, and $\sigma_{11}^{(\infty)} = \sigma_{33}^{(\infty)}$. Furthermore, $\sigma_{22}^{(\infty)} = \sigma_{44}^{(\infty)}$, due to the symmetry between the two systems. We  find the following quantities
\begin{align}
\tilde{\Delta}^{(\infty)} &=2  \,  \frac{ 4  \, \tilde{\alpha}^2+8 \, \tilde{ \alpha} +\tilde{\kappa}^2+4}{8  \, \tilde{\alpha} +\tilde{\kappa}^2+4} \, (2 \,  N_{\rm{th}}+1)^2\, , \nonumber \\
\det \boldsymbol{\sigma}^{(\infty)} &= \,  \frac{ 4 \, \tilde{\alpha}^2+8 \, \tilde{\alpha }+ \tilde{\kappa}^2+4}{8 \, \tilde{\alpha} + \tilde{\kappa}^2+4} \,(2 \,  N_{\rm{th}}+1)^4 \,  ,
\end{align}
from which it follows that 
\begin{equation}
(\tilde{\nu}_-^{(\infty)})^2  = \left(  2 \, N_{\rm{th}} + 1\right)^2   \left( \Lambda - \sqrt{ ( \Lambda - 1) \Lambda} \right) \, ,
\end{equation}
where $\Lambda = ( 4 \, \tilde{\alpha}^2 + 8 \, \tilde{\alpha} + \tilde{\kappa}^2 + 4)/(8 \, \tilde{\alpha} + \tilde{\kappa}^2 + 4)$.  We  note that the number of phonons $N_{\rm{th}}$ in the system has a strongly detrimental effect on the entanglement. For $E_N$ to be maximal, we require that $\nu_-$ is small. However, since $\tilde{\nu}_- \propto N_{\rm{th}}$, the entanglement decreases  as $N_{\rm{th}}$ increases. The decoherence rate $\tilde{\kappa}$, on the other hand, is not as influential as the phonon number. As $\tilde{\kappa}\rightarrow \infty$, we find that $\tilde{\nu}_- = |1 + 2 \, N_{\rm{th}}|$, while as $N_{\rm{th}} \rightarrow \infty$, we find $\nu_-^{(\infty)} \rightarrow \infty$. We therefore conclude that reducing the number of phonons in the system takes priority over reducing the decoherence rate. Consequently, the product $N_{\rm{th}} \tilde{\kappa}$ which is often quoted in experimental contexts is not as enlightening as an indicator of overall noise levels here, as the two quantities contribute differently to the entanglement. 

\begin{figure}[t!]
\subfloat[ \label{fig:density:alpha:vs:kappa}]{%
  \includegraphics[width=.45\linewidth, trim = -0.3cm 0cm 0.3cm 0cm]{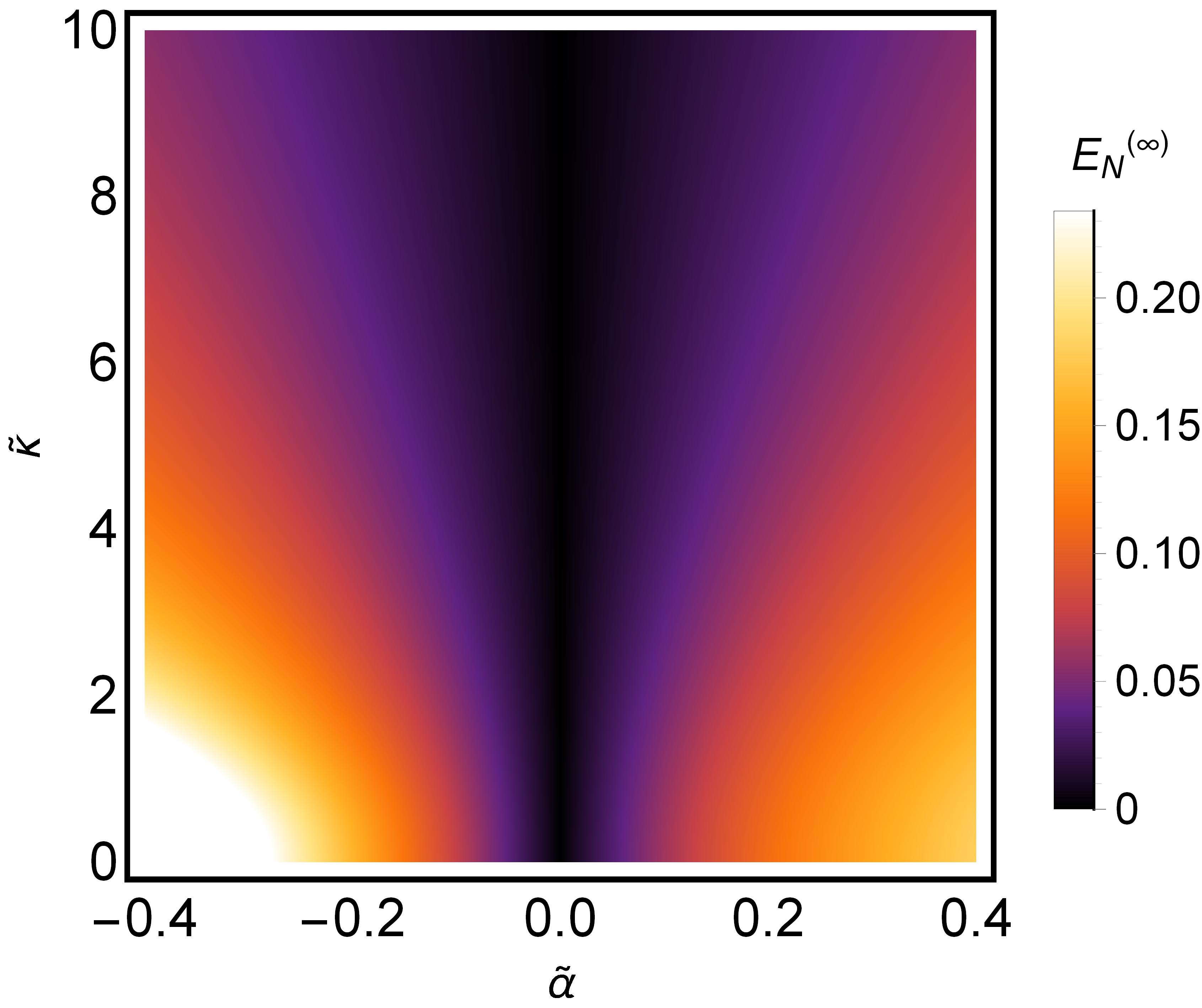}%
}\hfill $\quad$
\subfloat[  \label{fig:density:phonons:vs:kappa}]{%
  \includegraphics[width=.45\linewidth, trim =-0.3cm 0cm 0.3cm 0cm]{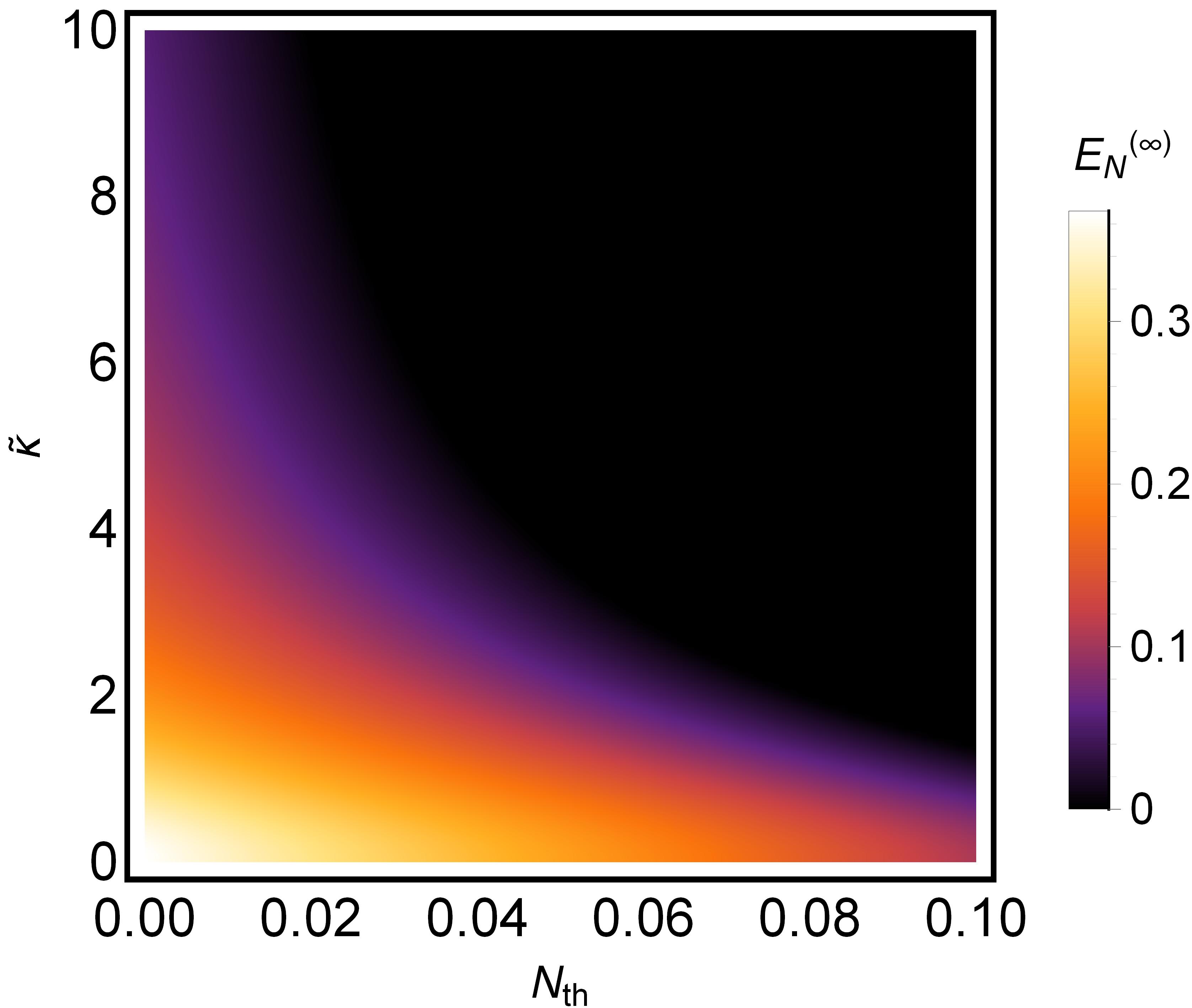}%
}\hfill
\caption{\small Steady state entanglement. \textbf{(a)} Density plot of the steady state entanglement $E_N^{(\infty)}$ as a function of  $\tilde{\alpha}$ and  $\tilde{\kappa}$ at $N_{\mathrm{th}} = 0$. The slight asymmetry is a feature of the interaction. \textbf{(b)} Density plot of the steady state entanglement $E_N^{(\infty)}$ as a function of $N_{\mathrm{th}}$ and $\tilde{\kappa}$ at $\tilde{\alpha} = -0.4$. }
\label{fig:noise:density}
\end{figure}

We further explore the entanglement in the steady state by plotting the logarithmic negativity $E_N^{(\infty)}$ for noisy dynamics in Fig.~\ref{fig:density:alpha:vs:kappa} for a range of interaction strengths $\tilde{\alpha} \in (-0.4, 0.4)$. As expected, when $\tilde{\kappa}$ increases, the logarithmic negativity $E_N^{(\infty)}$ decreases. The stronger the coupling, the more resilient the system is to noise.  Furthermore, all elements in Eq.~\eqref{eq:steady:state:covariance:elements} are proportional to the term $(2 N_{\mathrm{th}} + 1)$. To see how this affects the entanglement, we return to the PPT criterion. 
In Fig.~\ref{fig:density:phonons:vs:kappa}, we plot $E_N^{(\infty)}$ as a function of $N_{\rm{th}}$ and $\tilde{\kappa}$. The low values of $N_{\rm{th}}$ shown on the $x$-axis give an indication of how sensitive the system is to the number of thermal phonons in the system. 


\section{Discussion}
We have shown that entanglement from a central potential can be modelled to leading order with Gaussian states in the continuous variable framework. It remains to determine whether such entanglement can be detected in the laboratory.

\subsection{Error propagation}
To what precision must the entries in the covariance matrix $\boldsymbol{\sigma}$ be determined in order to verify the detection of entanglement? To estimate the required precision, we consider the asymptotic entanglement of the steady state $\boldsymbol{\sigma}^{(\infty)}$. We then assume that each element $\sigma_{ij}^{(\infty)}$ in the covariance matrix $\boldsymbol{\sigma}^{(\infty)}$ can be determined to within a specific precision $\epsilon_{ij}$, which we take to be a percentage of the total value of $\sigma_{ij}^{(\infty)}$, such that $\sigma_{11}^{(\infty)} \rightarrow \sigma_{11}^{(\infty)}(1 + \epsilon_{11})$. We also assume that the errors respect the symmetry of $\boldsymbol{\sigma}$ (so that, for example, $\epsilon_{21} = \epsilon_{12}$). From this assumption, we compute the error $\delta E_N^{(\infty)}$ via the standard error propagation formula:
\begin{equation} \label{eq:error:propagation:EN}
\delta E_N^{(\infty)} = \sqrt{\sum_k \left( \frac{\partial E_N^{(\infty)}}{\partial \epsilon_k } \epsilon_k \right)^2 }  \,  , 
\end{equation}
where in our case $k \in (11, 12, 13, 14, 22, 23, 24, 33, 34, 44)$.

\begin{figure*}
  \includegraphics[width=.49\linewidth, trim = 0cm 0cm 0cm 0cm]{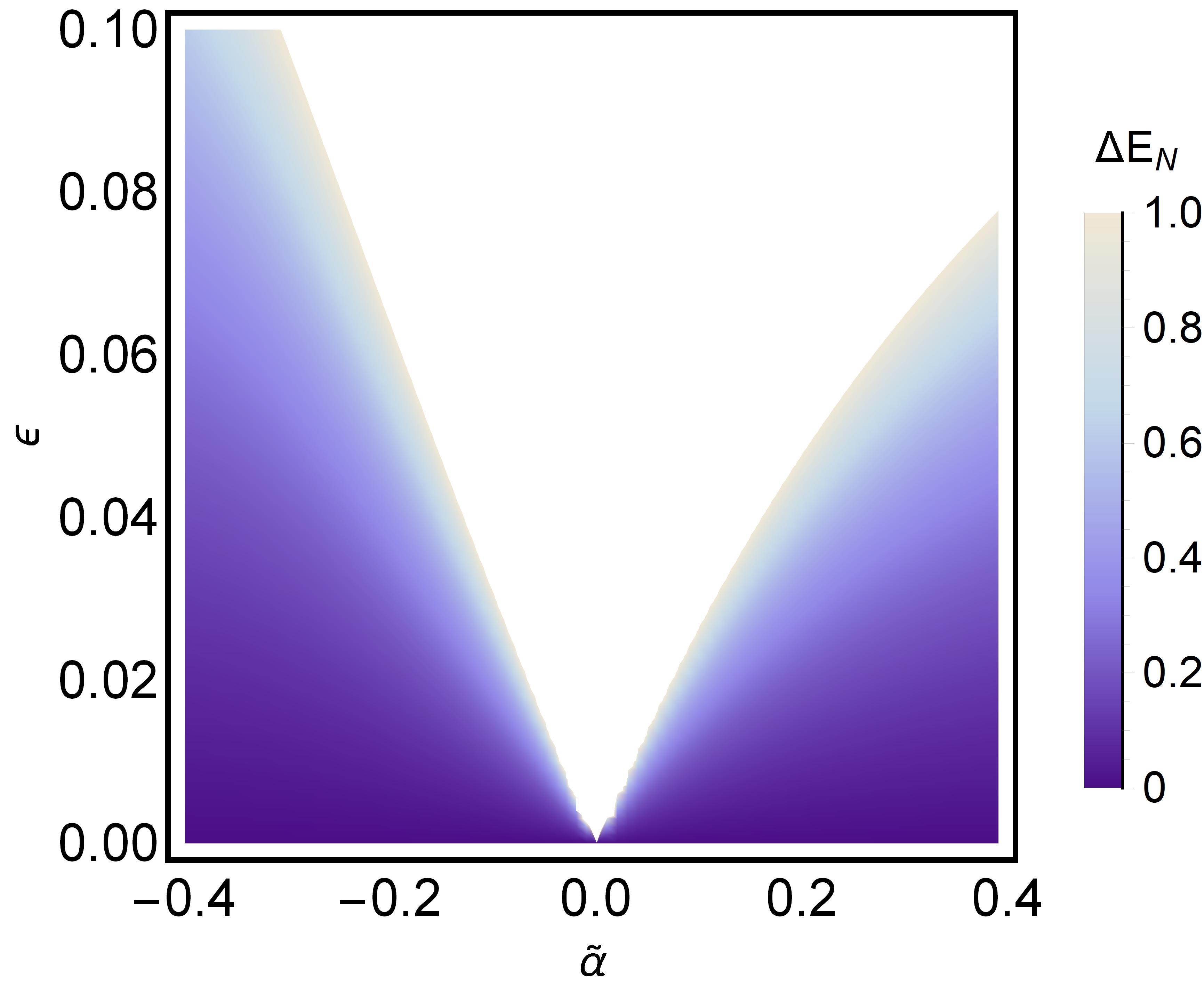}%
\hfill
\caption{\small Relative errors for detecting steady state entanglement.  Plot of the relative error $\Delta E_N = \delta E_N^{(\infty)}/E_N^{(\infty)}$ as a function of the coupling $\tilde{\alpha}$ and the precision $\epsilon$ at $\tilde{\kappa} = 1$ and $N_{\rm{th}} = 0$. For a weak coupling $|\tilde{\alpha}| \ll 1$, high precision is required to detect the entanglement. The asymmetry is due to the higher values of $E_N^{(\infty)}$ achieved from a negative coupling.  The plot does not show values for which the relative error is larger than unity. 
}
\label{fig:noise:density:error:EN}
\end{figure*}

To compute the precision required for entanglement detection, we make the sweeping assumption that all errors occur with the same magnitude, such that $\epsilon_{ij} \equiv \epsilon $ for all $i,j$. We then plot the relative error $\Delta E_N = \delta E_N^{(\infty)}/E_N^{(\infty)}$  in Figure~\ref{fig:noise:density:error:EN}. From the plots, it becomes evident that detecting logarithmic negativity from  small couplings $|\tilde{\alpha} | \ll 1$ requires a very small percentage error $\epsilon$.

\subsection{Entanglement through central potentials}

We now specialise to the following potentials: the Coulomb potential and  the Newtonian potential. Given our definition of $\tilde{\alpha}$ in Eq.~\eqref{eq:rescaled:coupling}, the couplings become, respectively:
\begin{align} \label{eq:potentials}
\tilde{\alpha}_{\mathrm{Cl}} &= \frac{q_1 q_2}{4 \pi \epsilon_0 \, r^3 \, m \omega_{\mathrm{m}}^2},
\quad\quad\quad \tilde{\alpha}_{\mathrm{Nw}} = - \frac{ G m }{r^3 \,  \omega_{\mathrm{m}}^2}, 
\end{align}
where  $q_1$ and $q_2$ are the charges on the bipartite system (where the signs will determine whether the potential is attractive or repulsive), $\epsilon_0$ is the vacuum permittivity, and $G$ is Newton's constant.

In what follows, we will consider state-of-the-art parameters for each case and determine whether entanglement due to this interaction can be realistically detected. This amounts to examining how close we can set $\tilde{\alpha}\sim -0.5$ for each interaction, which is the largest allowed value for an attractive potential. 

\subsubsection{Coulomb potential}
We consider two optomechanical spheres, each with a single but opposite charge $q_1 = -q_2 = e$, where $e$ is the electron charge, which results in an attractive potential with $\tilde{\alpha} <0$. The number of charges on each sphere can be controlled to exquisite precision by using ultraviolet light~\cite{moore2014search}. Linearised Coulomb interactions are commonly considered in trapped ions, and have also led to  the generation of entanglement~\cite{wilson2014tunable} but have  not yet been implemented for optomechanical systems, although a protocol to enhance cavity-mechanical entanglement through additional Coulomb interactions was proposed in~\cite{chen2015dissipation}. 

With the parameters in Table~\ref{tab:Coulomb}, it is possible to achieve a coupling $\tilde{\alpha}_{\mathrm{Cl}} \approx - 0.231$, meaning entanglement due to direct Coulombic interactions should be readily implementable for levitated systems. For a different choice of parameters, the attractive coupling can be made even stronger. We note that the mass-scaling is inverted, such that $\tilde{\alpha}_{\rm{Cl}} \propto m^{-1}$, which means  that entanglement due to the Coulomb potential is suppressed for systems with a larger mass. The number of thermal phonons in the system must still be low at $N_{\rm{th}} = 10^{-2}$ to allow for detection of entanglement, but the decoherence damping rate can be large at $\tilde{\kappa} = 1$ as it does not affect the system as much. If we require the error of the entanglement to be no more than $7\%$, we require that the quantities be determined to within $1\%$ of their value.

\def\arraystretch{1.2} 
\begin{table}[h!]
\caption{Values used to compute the entanglement from the attractive Coulomb potential.}
\begin{tabular}{l c c}
\multicolumn{3}{c}{\textbf{Coulomb potential}}\\ \hline 
\textbf{Parameter} & \textbf{Symbol} & \textbf{Value} \\
\hline
Mechanical frequency & $\omega_{\rm{m}}$ &  $100$ Rad\,s$^{-1}$  \\
\hline
Charge & $q_1, q_2$  & $1.602 \times 10^{-19}$  C \\
\hline
Oscillator mass & $m$  & $10^{-16}$ kg   \\
\hline
Separation & $r$ & $10^{-5}$ m \\
\hline
Coupling strength & $\tilde{\alpha}_{\mathrm{Cl}}
$ & $-0.231$ \\ \hline
Phonon decoherence & $\tilde{\kappa}$ & 1 \\ \hline
Thermal phonons & $N_{\rm{th}}$ & $10^{-2}$\\ \hline
Precision & $\epsilon$ & $10^{-2}$ \\ \hline
Steady state entanglement & $E_N^{(\infty)}$ & $0.119 \pm 0.008$ \\\hline
\end{tabular} \label{tab:Coulomb}
\end{table}


\subsubsection{Newtonian potential}

We first compute $\tilde{\alpha}$ for the parameters suggested in~\cite{bose2017spin} to see whether gravitational entanglement can be detected with only Gaussian states. 
With $m = 10^{-14}$ kg, $r = 200 \times 10^{-6}$ m and $\omega_{\mathrm{m}} = 10$ Hz, we find $\tilde{\alpha}_{\mathrm{Nw}} = -8.34 \times 10^{-16}$. Such a weak coupling will not yield any detectable entanglement within this scheme. 
Even if these figures were more lenient, there is the added complication that the Casimir--Polder interaction will typically dominate over gravity for most parameter choices. Nevertheless, we compute $\tilde{\alpha}_{\mathrm{Nw}}$ from the optimistic  parameters found in Table~\ref{tab:Newtonian}.  For these values, we still only find  $\tilde{\alpha}_{\mathrm{Nw}} = -6.67 \times10^{-8}$. For entanglement to be detectable at this interaction strength, we require that as few as $10^{-9}$ thermal phonons are in the system, but we find a slightly more forgiving  $\tilde{\kappa} = 0.1$. Finally, we require an extremely high precision of $\epsilon = 10^{-8}$ (which we recall is the percentage of the  covariance matrix elements).  

\begin{table}[h!]
\caption{ Values used to compute the entanglement from a Newtonian potential. }
\begin{tabular}{l c c}
\multicolumn{3}{c}{\textbf{Newtonian potential}}\\  \hline
\textbf{Parameter} & \textbf{Symbol} & \textbf{Value} \\
\hline
Mechanical frequency & $\omega_{\rm{m}}$ &  $10$ Rad\,s$^{-1}$  \\
\hline
Oscillator mass & $m$  & $10^{-13}$ kg   \\
\hline
Separation & $r$ & $10^{-6}$ m \\
\hline
Coupling strength & $\tilde{\alpha}_{\mathrm{Nw}}$ & $-6.67\times10^{-8}$ \\ \hline
Phonon decoherence & $\tilde{\kappa}$ & 0.1 \\ \hline
Thermal phonons & $N_{\rm{th}}$ & $10^{-9}$\\ \hline
Precision & $\epsilon$ & $10^{-8}$ \\ \hline
Steady state entanglement & $E_N^{(\infty)}$ & $(3.3 \pm 0.7)\times 10^{-8}$ \\\hline
\end{tabular} \label{tab:Newtonian}
\end{table}

For these values, we obtain an extremely small logarithmic negativity of $E_N^{(\infty)} = (3.3 \pm 0.7)\times 10^{-8} $, where the error stands at $22\%$.  Given these numbers, we conclude that gravitational entanglement is not likely to be detectable in the near term with the use of only Gaussian resources.

\subsection{Experimental detection}
A crucial step in the experimental generation of entanglement is the detection stage, where the state is measured to verify the entanglement. Such a detection scheme has, for example, been proposed in the context of a pulsed optomechanical setup~\cite{clarke2019generating}. Here, the light--matter interaction is confined to an optical pulse of a timescale much shorter than that of the mechanical oscillation period~\cite{vanner2011pulsed}, which means that it can be treated as a single unitary operator that entangles the light and mechanics. By then performing the appropriate measurements on the light, the mechanical state can be inferred by the pulse. 
In~\cite{clarke2019generating}, a measurement scheme is proposed to measure all first and second moments of two mechanical elements, which have been previously entangled through optical pulses correlated via the inclusion of a beamsplitter. The entangling step can be readily replaced by the central-potential entanglement scheme considered here.

\section{Conclusions}
In this work, we computed the leading-order entanglement due to a generic central-potential interaction between two levitated nanobeads. We derived the Hamiltonian matrix for a linearised potential and investigated the entanglement arising between two initially squeezed states  given unitary and noisy dynamics. Furthermore, we derived an analytic expression for the steady state of the system in the noisy setting and proposed a simple continuous-variable test for detecting the entanglement. With these tools, we computed the entanglement from an attractive Coulomb potential and the Newtonian potential. By considering errors that occur when determining the covariance matrix elements, we determined the measurement precision required in each scenario. 

Most importantly, we emphasise that this particular setup will not suffice for the detection of entanglement due to gravity.  Our results suggest that the inclusion of non-Gaussian resources may play a significant role here (they may, for example, explain the viability of the scheme with freely-falling systems envisioned in Ref~\cite{bose2017spin}, which relies on the creation of highly non-Gaussian initial states). To prepare two trapped mesoscopic systems in non-Gaussian states, one may utilise the nonlinear optomechanical interaction~\cite{qvarfort2019enhanced}, which couples the photon number to the position of the mechanical element through the addition of a cavity~\cite{aspelmeyer_2014}. Specific schemes for generating mechanical cat-states through this interaction have already been proposed by means of the nonlinear optomechanical evolution~\cite{bose1997preparation, bose1999scheme}, a simple optical interferometry setup~\cite{marshall2003towards} or within the pulsed optomechanical regime~\cite{clarke2018growing}. We further limited our investigation to Markovian environmental noise in this work, however non-Markovian noise-sources are expected to affect optomechanical resonators at low temperatures~\cite{groblacher2015observation}. It is therefore imperative to investigate the effect of non-Markovian noise on entanglement in this setting, which can be done with a non-Markovian master equation~\cite{vasile2010nonclassical}. Finally, to mitigate the small entanglement rates, one may also consider the addition of feedback techniques, which by themselves cannot generate entanglement, but which can serve to amplify already present interactions. These questions, and further scenarios that take into account the noise from the trapping laser will be considered in future work. 

After this work was first completed, the authors became aware of similar work by Krisanda \textit{et al}.~\cite{krisnanda2020observable}.

\section*{Acknowledgments} We thank Sahar Sahebdivan, Alexander D. Plato,  Marko Toro\v{s}, Anja Metelmann, Gavin Morley, Dennis R\"{a}tzel, Nathana\"{e}l Bullier, Peter F. Barker,  Alessio Belenchia, and Markus Aspelmeyer for fruitful discussions. SQ is supported by an EPSRC Doctoral Prize Fellowship. 

\section*{Bibliography}
\bibliographystyle{iopart-num}
\bibliography{bibliography}

\end{document}